\documentclass{css2016}

\usepackage{epsfig}

\newcommand{\PACS}{\MSC}

\title{
The observed spatial distribution of matter on scales ranging from 100 kpc to 1Gpc is inconsistent with the standard dark-matter- based cosmological models
}

\author{Pavel Kroupa$^{1,2}$\\
\vskip 2mm {\small
  $^1$
  Helmholtz Institut f\"ur Strahlen und Kernphysik\\
  Universit\"at Bonn, Nussallee 14--16, 53115 Bonn, Germany\\
    $^2$ Charles University in Prague, Faculty of Mathematics and
  Physics, Astronomical Institute, V Hole\v{s}ovi\v{c}k\'ach 2, CZ-180
  00 Praha 8, Czech Republic\\
pavel@astro.uni-bonn.de
}}

\abstract{The spatial arrangement of galaxies (of satellites on a
  scale of $100\,$kpc) as well as their three-dimensional distribution
  in galaxy groups such as the Local Group (on a scale of $1\,$Mpc),
  the distribution of galaxies in the nearby volume of galaxies (on a
  scale of $8\,$Mpc) and in the nearby Universe (on a scale of
  $1\,$Gpc) is considered. There is further evidence that the CMB
  shows irregularities and for anisotropic cosmic expansion.  The
  overall impression one obtains, given the best data we have, is
  matter to be arranged as not expected in the dark-matter based
  standard model of cosmology (SMoC). There appears to be too much
  structure, regularity and organisation. Dynamical friction on the
  dark matter halos is a strong direct test for the presence of dark
  matter particles, but this process does not appear to be operative
  in the real Universe.  This evidence suggests strongly that
  dynamically relevant dark matter does not exist and therefore
  cosmology remains largely not understood
  theoretically. More-accepted awareness of this case would by itself
  constitute a major advance in research providing fabulous
  opportunities for bright minds, and the observational data strongly
  suggest that gravitation must be effectively Milgromian,
  corresponding to a generalized Poisson equation in the classical
  limit. Thus, physical cosmology offers a significant historically
  relevant opportunity for ground-breaking work, at least for those
  daring to do so.}



\begin{document}

\maketitle

\section{Introduction}
The direct searches for dark matter particles, which the vast majority
of researchers believe dominate the matter density of the Universe,
have been coming up empty handed despite a huge effort to find these
particles with various elaborate, large and expensive experiments on
Earth and in space. But the astronomical evidence has already been
showing that dark matter particles cannot be there. This seems to be a
contradictory statement, because astronomical evidence has been used
to argue for the existence of dark matter particles which must be new
particles not contained in the standard model of particle physics
(SMoPP), which is otherwise a well tested theory.

The argument is as follows: If it is assumed that the Universe is
described by Einstein's field equation\footnote{As emphasized in
  \cite{Kroupa12a,Kroupa15b} this is an extrapolation by many orders
  of magnitude in scale and gradient of the potential of an
  empirically derived law, strictly valid {\it only} on the scale of
  the Solar-system. See footnote~\ref{footnote:banik} for an analogy.}
such that Newtonian gravitation is valid in the classical regime and
if all the matter was produced in the Big Bang, then the rate with
which structures form as cosmic time progresses, and also the motions
of stars and gas in the emerging galaxies when compared to
observations, shows conclusively that gravity must be stronger than
provided by the matter we know. One hypothesis is that much more
gravitating matter, that is dark matter which cannot interact
electromagnetically with normal matter and which is not described by
the standard model of particle physics (SMoPP), is required to yield,
roughly, the observed effects. Given this result, the researcher can
now assume this model (Einstein plus dark matter, lets call it the
null hypothesis) to be valid and perform detailed calculations of
galactic systems to further test the hypothesis. Additional
assumptions (inflation and dark energy) are also needed and together
comprise the dark-matter-based standard model of cosmology (SMoC). A
discussion of the current status of the SMoC can be found in
\cite{Bull16} and a critical discussion is also provided by
\cite{KS14}.

This model can then be tested on various astronomical data, as
outlined below. The argument followed here is to proceed testing the
SMoC using the relative spatial distribution and, when available, the
relative motions of galaxies. The tests then become very robust, that
is, do not depend on the details of baryonic physics, since the tests
apply largely to the presence of galaxies within their dark matter
halos. Thus, if dark matter halos exist, their spatial arrangements
relative to each other and their motions relative to each other are
being tested, rather than the detailed ``sub-grid'' properties of
individual galaxies. Baryonic processes then only play a role in
determining if a dark matter halo hosts a galaxy or not, and arguably,
dark matter halos more massive than $10^{9}\,M_\odot$ are understood
to host galaxies with initial mass $>10^7\,M_\odot$. This text is a
short summary pointing to the relevant literature, rather than
providing the detailed analysis of each problem. More detailed
discussions of these issues, which this text is also based on, are
available in \cite{Kroupa10, Kroupa12a,Kroupa12b,Kroupa15a,
  Kroupa15b}.

The analysis of the distribution of galaxies in the Local Group can be
split into two parts: the distribution of satellite galaxies
(Sec.~\ref{sec:100kpc}), the distribution of non-satellite galaxies
(Sec.~\ref{sec:1Mpc}). The distribution of galaxies in the Local
Volume (Sec.~\ref{sec:8Mpc}) and the variation of the mean matter
density in the Local Universe provide further tests, in particular
also of the Cosmological Principle. This question is addressed
independently by probing evidence for isotropic cosmic expansion
(Sec.~\ref{sec:disc}). A direct test for the presence of 
dynamically relevant dark matter particles is provided by observable
consequences of dynamical friction (Sec.~\ref{sec:dynfric}).

\section{The 100$\,$kpc scale}
\label{sec:100kpc}

It is by now well established that the satellite galaxies around the
Milky Way are highly significantly distributed anisotropically in a
rotational disk-like structure with radius of about 250$\,$kpc and
thickness of about $30\,$kpc \cite{Kroupa05, Metz08, Metz09,
  Pawlowski12a, Pawlowski16}. The Andromeda galaxy has a richer
population of satellite galaxies with perhaps a number of planar
structures \cite{Metz07}, but one planar structure which contains
50~per cent of all satellites is even more pronounced and thinner than
that of the Milky Way \cite{Ibata13}. Both disks-of-satellites are
correlated \cite{Pawlowski13}.  Other major galaxies show significant
evidence that such satellite planes are common \cite{Ibata14a,
  Ibata15a}. The dwarf spheroidal galaxies in the M81 group, which is
the nearest Local-Group analogue (distance about $3.6\,$Mpc), are in a
flattened distribution \cite{Chiboucas13} and the satellite galaxies
of Cen~A (distance about $3.66\,$Mpc) are likewise in a plane, wich is
perpendicular to the dust lane of Cen~A \cite{Mueller16}. The
observational results are thus rather clear: disks of satellites are
common, and in fact they seem to be the rule rather than the
exception. This is impossible to be the case in the SMoC.

It has been shown that just to find the one Milky Way satellite system
in a dark-matter universe is very unlikely. To have such structures
around many hosts, let alone that the Milky Way and Andromeda systems
are correlated, essentially leads to a combined probability of zero,
assuming the SMoC to be valid and the dwarf galaxies in the Local
Group to be in their own dark matter halos \cite{Kroupa12a,
  Pawlowski14a}. Basically, this single observational result falsifies
the SMoC, as has been realized early-on already \cite{Kroupa05}.

Claims that the disks of satellites can be accounted for readily
within the SMoC such that they do not constitute a serious problem
have been found to be flawed \cite{Ibata14,
  Pawlowski14a,Pawlowski15a,Ibata15a}.  SMoC simulations show that
rotating disks of satellites are as unlikely within pairs of dark
matter halos (resembling the pair Milky Way--Andromeda) as in isolated
halos \cite{Pawlowski14b}.

The physical reason for this discrepancy between observation and the
SMoC is that the SMoC necessitates {\it all} Milky-Way-type dark
matter halos to form from numerous stochastic mergers of smaller halos
such that the result is that the distribution of dark-matter-dominated
satellite galaxies is spheroidal. Although the dark matter sub-halos
fall-in from cosmic filaments, these have widths larger than the
virial radii of the dark matter halos, such that the infall of
satellite galaxies, even if being anisotropic to some degree, remains
in significant disagreement with the thin disks-of-satellites, since
the Rosetta-orbits phase-mix and shrink through dynamical friction
\cite{Metz09b, Pawlowski12b}.  Indeed, the observed positions and
velocities of those satellite galaxies for which such data exist show
that infall-solutions do not exist, because dynamical friction on the
extended dark matter halos is too efficient \cite{Angus11}.

The only known physical process which can lead to such rotating disk
of satellites is that the dSph satellite galaxies are ancient tidal
dwarf galaxies (TDGs). How such populations can form has been
demonstrated \cite{Wetzstein07,Pawlowski11}. Such low-mass dwarf
galaxies cannot capture significant amounts of dark matter and their
putative dark matter content would then be explainable by Milgromian
gravitation \cite{MM13a, MM13b, Dabringhausen16}.

For future tests, \cite{PK13,Pawlowski15b} predict the proper motions
of the Milky Way satellite galaxies based on the argument that they
need to orbit within the disk-of satellites as otherwise the chance of
having such a vast polar structure for randomly moving satellites
would be negligibly small. And, if dSph satellites are ancient TDGs,
then their number is expected to correlate with other indicators for
past galaxy--galaxy encounters, such as the bulge mass \cite{Kroupa10,
  CK16}. This opens a possibility for further testing this notion
(i.e. that dSph satellite galaxies are mostly if not exclusively old
TDGs) through observational campaigns using small (also amateur)
telescopes \cite{Javanmardi16a}.

\section{The 1$\,$Mpc scale}
\label{sec:1Mpc}

The dwarf galaxies in the Local Group which are not satellites are
distributed in a very organized manner, namely in two $\approx50\,$kpc
thin planes of about 1.5$\,$Mpc extension, whereby each is nearly
equidistant from the line joining the Milky Way and Andromeda
\cite{Pawlowski13}. These structures suggest the Milky Way and
Andromeda to be causally connected, and this poses an important
constraint on models of the formation of the Local Group. The
physically best-motivated cause for this entire structure, including
the correlated disks of satellite systems around the Milky Way and
Andromeda, is for the two major galaxies to have had an encounter
about $9-11\,$Gyr ago \cite{Zhao13, BZ16}. This is only possible if
they do not have dark matter halos, as they would otherwise have
merged by now (e.g. \cite{Barnes98} for similar cases). The structure
of the Local Group is indeed not explainable within the dark matter
framework.

The next group of galaxies beyond the Local Group is the M81 system at
a distance of about 3.6$\,$Mpc. Here we already do not have such good
three-dimensional coordinate information, but the system of dwarf
galaxies in it is known to be highly anisotropic \cite{Chiboucas13} as
noted in Sec.~\ref{sec:100kpc}. The highly significant anisotropy in
the Cen~A group \cite{Mueller16} at a distance of approximately
$3.66\,$Mpc has been noted above.

Furthermore, the major galaxies in the M81 group have been
encountering each other at least once, because the system is filled
with tidal HI gas. This provides crucial information on the existence
of dark matter halos because solutions do not seem to exist which
explain the matter distribution as well as the present-day positions
and line-of-sight velocities \cite{Thomson99,Yun99}.  Essentially, if
dark matter halos exist, then this system ought to have already
merged. The probability that all three major inner galaxies of the M81
system have just met in the very recent (less than $1\,$Gyr) past
after forming independently is remotely small \cite{Oehm16}.

The same argument applies to compact groups of galaxies
\cite{Sohn15}. Too many compact groups are observed with a constancy
in number density with redshift such that they appear to be largely
non-merging in contradiction to the expectation in the SMoC. That the
compact groups have just assembled in the past $1\,$Gyr of their
observation with the member galaxies having formed independently of
each other constitutes a negligible physical possibility, especially
given the large number of such systems.  The only viable physical
explanation for the existence of so many compact groups is that the
galaxies in them interact for many$\,$Gyr without merging. This is not
possible in a dark-matter-based cosmological model.

\section{The 8Mpc scale}
\label{sec:8Mpc}

Cosmological structure is evident in the galaxy distribution within
the Local Volume which is a sphere with a radius of about $8\,$Mpc
around the Local Group. This volume contains galaxies within the local
sheet and also the local void. At least two fundamental problems with
the observed distribution of galaxies have been emphasized
\cite{PN10}: (i) the local void is too empty and (ii) massive galaxies are
too far from the sheet within the outer regions of the void. Each
problem individually they describe as being about $1\,$per cent or
less probable within the SMoC, such that the combined probability that
the observed distribution can arise in the SMoC approaches
zero. Consequently, structure formation appears to have proceeded
differently to the SMoC.

\section{The 1Gpc scale}
\label{sec:1Gpc}

The Local Universe on a scale of about $1\,$Gpc around the Local Group
should have small fluctuations in the density of galaxy counts, but
within about $300\,$Mpc of us the density decreases significantly with
decreasing distance to about 50~per cent its global cosmological value
posing serious tension with the SMoC \cite{Karachentsev12, Keenan13,
  Boehringer15,Whitbourn16}. The under-density on a sale of $300\,$Mpc
and less is significantly more pronounced than allowed by the SMoC
(fig.~1 in \cite{Kroupa15b}). This has bearing on the deduced
acceleration of cosmic expansion because photons arriving from larger
distances are redshifted more than in a homogenous universe. This may
be partially or entirely responsible for the dark-energy effect
\cite{Wiltshire07,Buchert08}, and this needs to be studied further.

\section{The lack-of-dynamical friction and lack-of-merging problems}
\label{sec:dynfric}

It has been noted by \cite{Shankar14} that the observed galaxy
population does not support the profusion of mergers that are expected
in the SMoC such that these authors argue that dynamical friction must
be less efficient.  This is consistent with the deficit of galaxies
with bulges compared to SMoC predictions and the survival of pure disk
galaxies since $8\,$Gyr \cite{Weinzirl09,Kormendy10,Lorenzo14, SS16}
and with the absence of the evolution of the ratio of the co-moving
number density of the most massive galaxies relative to less-massive
galaxies \cite{Conselice12}.  The absence of an evolution of the
number density of elliptical galaxies \cite{Delgado10} and the lack of
recoiled super-massive black holes \cite{Lena14} are furthermore also
consistent with this general lack of evidence for mergers being an
important process in the evolution of galaxies, in contrast to the
expectations from the SMoC.

But this is only possible if the dynamical influence of dark matter is
much below that in the SMoC (which would by itself be a violation of
the SMoC), or if dark matter does not exist, as otherwise the massive
and expansive dark matter halos around each galaxy are dictated by the
theory. That is, it is not possible to arbitrarily reduce the process
of dynamical friction to fit the data but keeping the dark matter
halos as obtained from the SMoC.  Consistent with this problem is the
lack of merging already observed in the M81 group of galaxies and in
the compact galaxy groups (Sec.~\ref{sec:1Mpc}).

\section{Discussion}
\label{sec:disc}

Given the statements in the Introduction, it is apparent that in order
to save the model we have two possibilities: 

\begin{enumerate}

\item We can shrug the problems away by arguing that we {\it simply
    know the model to be right anyway}. Such a statement is rather
  popular and is based on the main-stream understanding that the CMB
  is extremely well represented by the SMoC as evident with the Planck
  results. Any irregularities in the non-linear structure-formation
  regime (galaxy clusters, galaxies) are then not to be taken too
  seriously. But there are tensions between the CMB and the SMoC (see
  sec.~17.3.1 in \cite{Kroupa12a} and also \cite{Javanmardi15} and
  e.g. \cite{Grandis16}).  For example, there is tentative evidence
  for an unexpected alignment of various independent measures of
  anisotropy in the CMB, SN1a-based cosmological expansion and galaxy
  morphology possibly raising questions concerning the Cosmological
  Principle \cite{Javanmardi15, Javanmardi16}. There is also tension
  between the locally-measured Hubble constant and the Hubble constant
  as derived from the CMB \cite{Riess16,Beaton16}.  Ignoring such
  tensions and claiming excellent fits of the SMoC to the CMB as
  proving the dark-matter models to be correct may serve the
  short-term aims of a famous-few but undermines the very principle of
  natural scientific research, as such claims are based on belief
  rather than comprehensive evidence, remembering that no theory can
  ever be proven, but merely tested and if necessary discarded.

Thus, this avenue of thinking is not convincing.

\item It may be speculated that baryonic physics, which is described
  by perhaps the best model of physics we have (the SMoPP), conspires
  on every studied scale to annul the discrepancies in the sense that
  what we observed does not seem to match, but what we cannot see is
  an excellent account of reality.

Such an argument rests on speculation of unverifiable processes and
needs to be discarded.

\item The more scientific approach is to accept the failures and to
  seek an entirely different model. Such a model would need to be
  dark-matter free in order to test if baryonic structures alone,
  which are described by the best model of physics we currently have
  (the SMoPP), may be able to account for the observations, but in a
  different gravitational framework. Gravitation remains the least
  well understood force, if it is a force at all, and thus this ansatz
  appears to be the most promising avenue. Our work in Bonn and
  Strasbourg, using the Phantom of Ramses (PoR) computer code
  (\cite{Lueghausen15}, see also \cite{Candlish15}), developed with
  sparse funding from Bonn, is now allowing us to perform exactly this
  work in the Milgromian-gravitational framework
  \cite{Milgrom83,Milgrom09,Famaey12}.  The results so far appear
  highly promising \cite{RFK16, TKF16}.

\end{enumerate}

\section{Conclusion}
\label{sec:conc}

The above discussion suggests that the real Universe appears to
produce more structure, which is at least partially more ordered and
organized than the SMoC, and that the observed galaxy population
neither matches nor does it evolve as expected by the SMoC. The
explicit tests for the presence of dark matter via dynamical friction
suggest this process not to be acting. All of this is consistent with
the null results in the searches for dark matter particles. Here I
would wish to emphasize the incredible {\it consistency} of the tests
amongst each other: none of the tests performed yield positive results
concerning the SMoC, and all appear to suggest more structure and
organisation. This is important to note because we do not have the
situation where a test yields excellent agreement while another one
does not. They are all consistently problematical for the SMoC. In
\cite{Kroupa12a,Kroupa15a,Kroupa15b} the {\it theory confidence graph}
lists the many individual tests performed such that, if each failed
test decreases the confidence by 50~per cent then the remaining
confidence in the SMoC remains today at less than $10^{-5}\,$per
cent. 

We are left with inferring that the important hypothesis that dark
matter particles exist needs to be rejected by astronomical
data. Gravitation must therefore be
effectively\footnote{\label{footnote:banik}{\it Effectively}, because
  it may still be Einsteinian but with additional but non-exotic
  physics possibly playing a role in Minkowski space \cite{Milgrom99}.
  This is nicely visualized by an analogy by Indranil Banik: consider
  a trampoline. One can measure its depth-extension $s=s(w)$ as a
  function of weight $w$. These measurements can be fitted by an
  empirical law for macroscopic weights (e.g. $w>1\,$kg).  We would
  then not expect this same law $s=s(w)$ to hold in an extrapolation
  to $w<10^{-5}\,$g, for instance, because molecular forces will begin
  to play a role for very small $w$.} stronger on scales relevant for
galaxies. Mordehai Milgrom \cite{Milgrom83} has conceptualized a
generalized gravity known as MOND, or as Milgromian gravitation. This
finding can be seen as constituting the greatest advance in
gravitational physics since Newton and Einstein and it is based on a
generalized Poisson equation and a Lagrangian \cite{BM84} and can also
be embedded in a general-relativistic theory, as discovered by Jacob
Bekenstein \cite{Bekenstein04} with notable reviews
\cite{Bekenstein06, Bekenstein09} with alternatives \cite{BH15,
  Trippe15, Khoury16}. The observed deviations from Newtonian
gravitation at the very weak accelerations, which are described by
Milgromian gravity, may be a result of vacuum processes, perhaps as
discussed for Minkowski space by Milgrom \cite{Milgrom99}.  Milgromian
dynamics has proven to be extraordinarily successful \cite{Famaey12}
and is now being used in numerical experiments to study galaxy
formation and evolution \cite{Lueghausen15,TKF16,RFK16}. These
numerical experiments appear to be showing an incredible amount of
success in reproducing all major issues in the astrophysics of
galaxies, as our work at the Universities of Strasbourg and Bonn is
demonstrating. Further work will be published in due course, subject
to the availability of funding.

As to the issue of a more structured universe which may also be more
organized \cite{Llinares08}, it appears that a Milgromian universe may
provide the former, and self-regulatory growth processes may provide
the latter which may be related to the fundamental assumption of
conservation of matter.

Closing this critical discussion, one of the currently most fundamental
problems in theoretical physics is the origin of Milgromian dynamics
rather the nature of (non-existing) dark matter particles. This is
likely to be an immense opportunity for talented young
researchers. Concerning the theory of galactic astrophysics,
understanding the formation and evolution of galaxies in Milgromian
gravity provides a great opportunity for talented young researchers
interested in performing numerical astrophysics experiments.

\section*{Acknowledgements}
I thank the organizers of this conference in the ancient and
scientifically highly relevant city of Prague in September 2016 for
inviting me to provide this lecture.


\begin{thebibliography}{1}


\bibitem{Angus11} Angus, G.~W., Diaferio, A., \& Kroupa, P.: Using
dwarf satellite proper motions to determine their origin.  Monthly
Notices of the Royal Astronomical Society (2011) \textbf{416},
1401--1409.


\bibitem{BZ16} Banik, I., \& Zhao, H.: Dynamical history of the Local
  Group in $\Lambda$CDM.  Monthly Notices of the Royal Astronomical Society
  \textbf{459} (2016), 2237--2261.

\bibitem{Barnes98} Barnes, J.~E.: Dynamics of
  Galaxy Interactions. In: R. C. Kennicutt, Jr. F. Schweizer,
  J. E. Barnes, D. Friedli, L. Martinet, and D. Pfenniger (Eds.),
  \emph{ Galaxies: Interactions and Induced Star Formation, Saas-Fee
    Advanced Course 26. Lecture Notes 1996. Swiss Society for
    Astrophysics and Astronomy, XIV}, Springer-Verlag
  Berlin/Heidelberg; ISBN: 3-540-63569-6, 1998., p.275.


\bibitem{Beaton16} Beaton, R.~L., Freedman, W.~L., Madore, B.~F., et
  al.: The Carnegie-Chicago Hubble Program. I. A New Approach to the
  Distance Ladder Using Only Distance Indicators of Population II.
  Astronomical Journal, submitted (2016), arXiv:1604.01788.


\bibitem{Bekenstein04} Bekenstein, J.~D.:
  Relativistic gravitation theory for the modified Newtonian dynamics
  paradigm.  Physical Review D \textbf{70} (2004), 083509.

\bibitem{Bekenstein06} Bekenstein, J.: The modified Newtonian dynamics
  - MOND and its implications for new physics.  Contemporary Physics
  \textbf{47} (2006), 387--403

\bibitem{Bekenstein09} Bekenstein, J.~D.: Relativistic MOND as an
  alternative to the dark matter paradigm. Nuclear Physics A
  \textbf{827} (2009), 555--560.

\bibitem{BM84} Bekenstein, J., \& Milgrom, M.: Does the missing mass
  problem signal the breakdown of Newtonian gravity?  The
  Astrophysical Journal \textbf{286} (1984), 7--14.


\bibitem{BH15} Blanchet, L., \& Heisenberg, L.: Dipolar dark matter
  with massive bigravity.  Journal of Cosmology and Astroparticle
  Physics \textbf{12} (2015), 026



\bibitem{Boehringer15} B{\"o}hringer, H., Chon, G., Bristow, M., \&
  Collins, C.~A.: The extended ROSAT-ESO Flux-Limited X-ray Galaxy
  Cluster Survey (REFLEX II). V. Exploring a local underdensity in the
  southern sky. Astronomy \& Astrophysics \textbf{574} (2015), 26--34.



\bibitem{Buchert08} Buchert, T.: Dark Energy from structure: a status
  report.  General Relativity and Gravitation \textbf{40} (2008),
  467--527.


\bibitem{Bull16} Bull, P., Akrami, Y., Adamek, J., et al.: Beyond
  $\Lambda$CDM: Problems, solutions, and the road ahead.  Physics of
  the Dark Universe \textbf{12} (2016), 56--99.


\bibitem{Candlish15} Candlish, G.~N., Smith, R., \& Fellhauer, M.:
  RAyMOND: an N-body and hydrodynamics code for MOND.  Monthly Notices
  of the Royal Astronomical Society \textbf{446} (2015), 1060--1070.


\bibitem{Conselice12} Conselice, C.~J.: Galaxy Formation: Where Do We
  Stand?  In: VIII International Workshop on the Dark Side of the
  Universe, 2012, arXiv:1212.5641.

\bibitem{Chiboucas13} Chiboucas, K., Jacobs, B.~A., Tully, R.~B., \&
  Karachentsev, I.~D.: Confirmation of Faint Dwarf Galaxies in the M81
  Group.  The Astronomical Journal \textbf{146} (2013), 126--160.


\bibitem{Dabringhausen16} Dabringhausen, J., Kroupa, P., Famaey, B.,
  Fellhauer M.: Understanding the internal dynamics of elliptical
  galaxies without non-baryonic dark matter. Monthly Notices of the
  Royal Astronomical Society (2016), accepted.


\bibitem{Delgado10} Delgado-Serrano, R., Hammer, F., Yang, Y.~B., et
  al.: How was the Hubble sequence 6 Gyr ago?  Astronomy \&
  Astrophysics \textbf{509} (2010), 78--.


\bibitem{Famaey12} Famaey, B., \& McGaugh, S.~S.: Modified Newtonian
  Dynamics (MOND): Observational Phenomenology and Relativistic
  Extensions.  Famey12Living Reviews in Relativity \textbf{15} (2012)


\bibitem{Lorenzo14} Fern{\'a}ndez Lorenzo, M., Sulentic, J.,
  Verdes-Montenegro, L., et al.: Are (Pseudo)bulges in Isolated
  Galaxies Actually Primordial Relics?  The Astrophysical Journal
  Letters \textbf{788} (2014), L39--L45.

\bibitem{Grandis16} Grandis, S., Rapetti, D., Saro, A., Mohr, J.~J.,
  \& Dietrich, J.~P.: Quantifying Tensions between CMB and Distance
  Datasets in Models with Free Curvature or Lensing Amplitude.
  (2016), arXiv:1604.06463

\bibitem{Ibata13} Ibata, R.~A., Lewis, G.~F., Conn, A.~R., et al.: A
  vast, thin plane of corotating dwarf galaxies orbiting the Andromeda
  galaxy. Nature \textbf{493} (2013), 62--65.

\bibitem{Ibata14a} Ibata, N.~G., Ibata, R.~A., Famaey, B., \& Lewis,
  G.~F.: Velocity anti-correlation of diametrically opposed galaxy
  satellites in the low-redshift Universe. Nature \textbf{511} (2014),
  563--566.

\bibitem{Ibata14} Ibata, R.~A., Ibata, N.~G., Lewis, G.~F., et al.: A
  Thousand Shadows of Andromeda: Rotating Planes of Satellites in the
  Millennium-II Cosmological Simulation. The Astrophysical Journal
  Letters \textbf{784} (2014), L6--L11.

\bibitem{Ibata15a} Ibata, R.~A., Famaey, B., Lewis, G.~F., Ibata,
  N.~G., \& Martin, N.: Eppur si Muove: Positional and Kinematic
  Correlations of Satellite Pairs in the Low Z Universe.  The
  Astrophysical Journal \textbf{805} (2015), 67--77.


\bibitem{Javanmardi15} Javanmardi, B., Porciani, C., Kroupa, P., \&
  Pflamm-Altenburg, J.: Probing the Isotropy of Cosmic Acceleration
  Traced By Type Ia Supernovae.  The Astrophysical Journal
  \textbf{810} (2015), 47--57.

\bibitem{Javanmardi16a} Javanmardi, B., Martinez-Delgado, D., Kroupa,
  P., et al.: DGSAT: Dwarf Galaxy Survey with Amateur
  Telescopes. I. Discovery of low surface brightness systems around
  nearby spiral galaxies.  Astronomy \& Astrophysics \textbf{588}
  (2016), 89--101.


\bibitem{Javanmardi16} Javanmardi, B., \& Kroupa, P.: Anisotropy in
  the all-sky distribution of galaxy morphologies.  Astronomy \&
  Astrophysics, in press, arXiv:astro-ph/1609.06719 


\bibitem{Karachentsev12} Karachentsev, I.~D.: Missing dark matter in
  the local universe.  Astrophysical Bulletin \textbf{67} (2012), 123--134.


\bibitem{Keenan13} Keenan, R.~C., Barger, A.~J., \& Cowie, L.~L.:
  Evidence for a ~300 Megaparsec Scale Under-density in the Local
  Galaxy Distribution.  The Astrophysical Journal \textbf{775} (2013),
  62--78.


\bibitem{Khoury16} Khoury, J.: Another path for the emergence of
  modified galactic dynamics from dark matter superfluidity.  Physical
  Review D \textbf{93} (2016), 103533.


\bibitem{Kormendy10} Kormendy, J., Drory, N., Bender, R., \& Cornell,
  M.~E.: Bulgeless Giant Galaxies Challenge Our Picture of Galaxy
  Formation by Hierarchical Clustering.  The Astrophysical Journal
  \textbf{723} (2010), 54--80.

\bibitem{KS14} Krizek, M., \& Somer, L., A critique of the standard
  model of cosmology. Neural Network World \textbf{24} (2014), 435--.

\bibitem{Kroupa05} Kroupa, P., Theis, C., \& Boily, C.~M.: The great
  disk of Milky-Way satellites and cosmological
  sub-structures. Astronomy \& Astrophysics \textbf{431} (2005),
  517--521.


\bibitem{Kroupa10} Kroupa, P., Famaey, B., de Boer, K.~S., et al.:
  Local-Group tests of dark-matter concordance cosmology. Towards a
  new paradigm for structure formation.  Astronomy \& Astrophysics
  \textbf{523} (2010), 32--54.

\bibitem{Kroupa12a} Kroupa, P.: The Dark Matter Crisis:
  Falsification of the Current Standard Model of
  Cosmology. Publications of the Astronomical Society of Australia
  \textbf{29} (2012), 395--433.

\bibitem{Kroupa12b} Kroupa, P., Pawlowski, M., \& Milgrom, M.: The
Failures of the Standard Model of Cosmology Require a New Paradigm. 
International Journal of Modern Physics D \textbf{21} (2012), 1230003 

\bibitem{Kroupa15a}: Lessons from the Local Group (and beyond) on dark
  matter. In: K. C. Freeman, B. G. Elmegreen, D. L. Block, and
  M. Woolway (Eds.), \emph{Lessons from the Local Group} , Dordrecht:
  Springer, 2015.


\bibitem{Kroupa15b} Kroupa, P.: Galaxies as simple dynamical systems:
  observational data disfavor dark matter and stochastic star
  formation. Canadian Journal of Physics \textbf{93} (2015), 169--202.


\bibitem{Lena14} Lena, D., Robinson, A., Marconi, A., et al.:
  Recoiling Supermassive Black Holes: A Search in the Nearby
  Universe. The Astrophysical Journal \textbf{795} (2014), 146--177.

\bibitem{Llinares08} Llinares, C., Knebe, A., \& Zhao, H.:
  Cosmological structure formation under MOND: a new numerical solver
  for Poisson's equation.  Monthly Notices of the Royal Astronomical
  Society \textbf{391} (2008), 1778--1790


\bibitem{Lueghausen15} L{\"u}ghausen, F., Famaey, B., \& Kroupa, P.:
Phantom of RAMSES (POR): A new Milgromian dynamicsN-body code.
Canadian Journal of Physics \textbf{93} (2015), 232--241.



\bibitem{CK16} L{\'o}pez-Corredoira, M., \& Kroupa, P.: The Number of
  Tidal Dwarf Satellite Galaxies in Dependence of Bulge Index.  The
  Astrophysical Journal \textbf{817} (2016), 75--82.


\bibitem{MM13a} McGaugh, S., \& Milgrom, M.: Andromeda Dwarfs in Light
  of Modified Newtonian Dynamics.  The Astrophysical Journal
  \textbf{766} (2013), 22--29.

\bibitem{MM13b} McGaugh, S., \& Milgrom, M.: Andromeda Dwarfs in Light
  of Modified Newtonian Dynamics.  The Astrophysical Journal
  \textbf{775} (2013), 139--145.

\bibitem{Metz07} Metz, M., Kroupa, P., \& Jerjen, H.: The spatial
  distribution of the Milky Way and Andromeda satellite galaxies.
  Monthly Notices of the Royal Astronomical Society \textbf{374}
  (2007), 1125--1145.

\bibitem{Metz08} Metz, M., Kroupa, P.,
  \& Libeskind, N.~I.: The Orbital Poles of Milky Way Satellite
  Galaxies: A Rotationally Supported Disk of Satellites. The
  Astrophysical Journal \textbf{680} (2008), 287--294. 

\bibitem{Metz09} Metz, M., Kroupa, P., \& Jerjen, H.: Discs of
  satellites: the new dwarf spheroidals. Monthly Notices of the Royal
  Astronomical Society \textbf{394} (2009), 2223--2228.

\bibitem{Metz09b} Metz, M., Kroupa, P., Theis, C., Hensler, G., \&
  Jerjen, H.: Did the Milky Way Dwarf Satellites Enter The Halo as a
  Group?  The Astrophysical Journal \textbf{697} (2009), 269--274.

\bibitem{Milgrom83} Milgrom, M.: A modification of the Newtonian
  dynamics as a possible alternative to the hidden mass hypothesis.
  The Astrophysical Journal \textbf{270} (1983), 365--370.


\bibitem{Milgrom99} Milgrom, M.: The modified dynamics as a vacuum
  effect.  Physics Letters A \textbf{253} (1999), 273--279.


\bibitem{Milgrom09} Milgrom, M.: The Mond Limit from Spacetime Scale
  Invariance.  The Astrophysical Journal \textbf{698} (2009),
  1630--1638.



\bibitem{Mueller16} M{\"u}ller, O., Jerjen, H., Pawlowski, M.~S., \&
  Binggeli, B.: Testing the two planes of satellites in the Centaurus
  Group. Astronomy \& Astrophysics (2016), submitted, arXiv:1607.04024

\bibitem{Oehm16} Oehm, W., Thies, I., Kroupa, P.: in prep.

\bibitem{Pawlowski11} Pawlowski, M.~S., Kroupa, P., \& de Boer, K.~S.:
  Making counter-orbiting tidal debris. The origin of the Milky Way
  disc of satellites? Astronomy \& Astrophysics \textbf{532} (2011),
  118--143.

\bibitem{Pawlowski12a} Pawlowski, M.~S., Pflamm-Altenburg, J., \&
  Kroupa, P.: The VPOS: a vast polar structure of satellite galaxies,
  globular clusters and streams around the Milky Way.  Monthly Notices
  of the Royal Astronomical Society \textbf{423} (2012), 1109--1126.


\bibitem{Pawlowski12b} Pawlowski, M.~S., Kroupa, P., Angus, G., et al.:
  Filamentary accretion cannot explain the orbital poles of the Milky
  Way satellites. Monthly Notices of the Royal Astronomical Society
  \textbf{424} (2012), 80--92.

\bibitem{Pawlowski13} Pawlowski, M.~S., Kroupa, P., \& Jerjen, H.:
  Dwarf galaxy planes: the discovery of symmetric structures in the
  Local Group.  Monthly Notices of the Royal Astronomical Society
  \textbf{435} (2013), 1928--1957.


\bibitem{PK13} Pawlowski, M.~S., \& Kroupa, P.: The rotationally
  stabilized VPOS and predicted proper motions of the Milky Way
  satellite galaxies. Monthly Notices of the Royal Astronomical
  Society \textbf{435} (2013), 2116--2131.

\bibitem{Pawlowski14a} Pawlowski, M.~S., Famaey, B., Jerjen, H., et
  al.: Co-orbiting satellite galaxy structures are still in conflict
  with the distribution of primordial dwarf galaxies.  Monthly Notices
  of the Royal Astronomical Society \textbf{442} (2014), 2362--2380.


\bibitem{Pawlowski14b} Pawlowski, M.~S., \& McGaugh, S.~S.:
  Co-orbiting Planes of Sub-halos are Similarly Unlikely around Paired
  and Isolated Hosts. The Astrophysical Journal Letters \textbf{789}
  (2014), L24--L31.




  \bibitem{Pawlowski15a} Pawlowski, M.~S., Famaey, B., Merritt, D., \&
  Kroupa, P.: On the Persistence of Two Small-scale Problems in
  ¦«CDM. The Astrophysical Journal \textbf{815} (2015), 19--31.

\bibitem{Pawlowski15b} Pawlowski, M.~S., McGaugh, S.~S., \& Jerjen,
  H.: The new Milky Way satellites: alignment with the VPOS and
  predictions for proper motions and velocity dispersions.  Monthly
  Notices of the Royal Astronomical Society \textbf{453} (2015),
  1047--1061.


\bibitem{Pawlowski16} Pawlowski, M.~S.: The alignment of SDSS
  satellites with the VPOS: effects of the survey footprint
  shape. Monthly Notices of the Royal Astronomical Society
  \textbf{456} (2016), 448--458.

\bibitem{PN10} Peebles, P.~J.~E., \& Nusser, A.: Nearby galaxies as
  pointers to a better theory of cosmic evolution. Nature \textbf{465}
  (2010), 565--569.

\bibitem{RFK16} Renaud, F., Famaey, B., \& Kroupa P.
Star formation triggered by galaxy interactions in Milgromian gravity.
Monthly Notices of the Royal Astronomical society, submitted.


\bibitem{Riess16} Riess, A.~G., Macri, L.~M., Hoffmann, S.~L., et al.:
  A 2.4\% Determination of the Local Value of the Hubble Constant.
  The Astrophysical Journal \textbf{826} (2016), 56-- .


\bibitem{SS16} Sachdeva, S., \& Saha, K.: Survival of Pure Disk
  Galaxies over the Last 8 Billion Years.  The Astrophysical Journal
  Letters \textbf{820} (2016), L4--L9.


\bibitem{Shankar14} Shankar, F., Mei, S., Huertas-Company, M., et al.:
  Environmental dependence of bulge-dominated galaxy sizes in
  hierarchical models of galaxy formation. Comparison with the local
  Universe. Monthly Notices of the Royal Astronomical Society
  \textbf{439} (2014), 3189--3212.


\bibitem{Sohn15} Sohn, J., Hwang, H.~S., Geller, M.~J., et al.:
  Compact Groups of Galaxies with Complete Spectroscopic Redshifts in
  the Local Universe.  Journal of Korean Astronomical Society
  \textbf{48} (2015), 381--398.

\bibitem{TKF16} Thies, I., Kroupa, P., \& Famaey, B.: Simulating disk
  galaxies and interactions in Milgromian dynamics.  In: Second
  BELISSIMA Workshop, arXiv:1606.04942, 2016




\bibitem{Thomson99} Thomson, R.~C., Laine, S., \& Turnbull, A.:
  Towards an Interaction Model of M81, M82 and NGC 3077. In:
  J. E. Barnes, and D. B. Sanders (Eds), \emph{ Galaxy Interactions at
    Low and High Redshift}, IAU Symposium Nr. 186, p.135, 1999.



\bibitem{Trippe15} Trippe, S.: Milgrom's Law and Λ's Shadow: How
  Massive Gravity Connects Galactic and Cosmic Dynamics.  Journal of
  Korean Astronomical Society \textbf{48} (2015), 191--194.


\bibitem{Weinzirl09} Weinzirl, T., Jogee, S., Khochfar, S., Burkert,
  A., \& Kormendy, J.: Bulge n and B/T in High-Mass Galaxies:
  Constraints on the Origin of Bulges in Hierarchical Models.  The
  Astrophysical Journal \textbf{696} (2009), 411--447.

\bibitem{Wetzstein07} Wetzstein, M., Naab, T., \& Burkert, A.: Do
  dwarf galaxies form in tidal tails? Monthly Notices of the Royal
  Astronomical Society \textbf{375} (2007), 805--820.


\bibitem{Whitbourn16} Whitbourn, J.~R., \& Shanks, T.: The galaxy
  luminosity function and the Local Hole.  Monthly Notices of the
  Royal Astronomical Society \textbf{459} (2016), 496--507.


\bibitem{Wiltshire07} Wiltshire, D.~L.: Cosmic clocks, cosmic variance
  and cosmic averages.  New Journal of Physics \textbf{9} (2007), 377--.


\bibitem{Yun99} Yun, M.~S.: Tidal Interactions in M81 Group. In:
  J. E. Barnes, and D. B. Sanders (Eds), \emph{ Galaxy Interactions at
    Low and High Redshift}, IAU Symposium Nr. 186, p.81, 1999.


\bibitem{Zhao13} Zhao, H., Famaey, B., L{\"u}ghausen, F., \& Kroupa,
  P.: Local Group timing in Milgromian dynamics. A past Milky
  Way-Andromeda encounter at $z > 0.8$. Astronomy \& Astrophysics
  \textbf{557} (2013), L3--L7.

\end{thebibliography}
\end{document}